\documentclass[12pt]{article}
\usepackage{amsfonts}
\usepackage{amsmath}
\begin{document}
\title{Chow's theorem and universal holonomic quantum computation}
\author{Dennis Lucarelli \\ Department of Systems Science and Mathematics \\
Washington University, St. Louis, Missouri 63130}
\date{}
\maketitle
\begin{abstract} A theorem from control theory relating the Lie algebra generated by vector
fields on a manifold to the controllability of the dynamical
system is shown to apply to Holonomic Quantum Computation.
Conditions for deriving the holonomy algebra are presented by
taking covariant derivatives of the curvature associated to a
non-Abelian gauge connection.  When applied to the Optical
Holonomic Computer, these conditions determine that the holonomy
group of the two-qubit interaction model contains $SU(2) \times
SU(2)$.  In particular, a universal two-qubit logic gate is
attainable for this model.

\vspace{10pt}
\end{abstract}
\section{Introduction}
Controlling quantum dynamics to effect a desired unitary evolution
is a fundamental issue in quantum computation.  Full control over
the system dynamics and hence the ability to realize any logic
gate is called \emph{universal quantum computation}.  In recent
years, there has been considerable interest in conditions for
universality and it has been proven that for an $n$-level quantum
system universality is a \emph{dense} condition, being satisfied
by \emph{almost all} computational models
\cite{deutsch}-\cite{nik}.

Despite the richness of the mathematical model,
the effects of quantum  noise make it a tremendous challenge to manifest this property in
nature.  Zanardi and Rasetti \cite{zan} have proposed a  novel methodology for
the control of quantum information which may provide a resolution to these competing
phenomena.  Holonomic Quantum Computation (HQC), as introduced in \cite{zan},
is a theoretically appealing model that can provide universal computation,
and, due to the geometrical  nature of the framework, possesses intrinsic
robustness against decoherence and control imperfection.

In this paper, we are primarily interested in the existence of the logic gates
available to the experimentalist within the HQC framework.  Specifically, we report a result
from geometric control theory that simplifies the calculation of the holonomy group associated to
a non-Abelian gauge connection.  The application of this theorem to HQC establishes conditions
for universality in general and in particular proves the universality of the
\emph{Optical Holonomic Computer} \cite{zan}-\cite{pach2}.
We refer the reader to the literature for a more detailed exposition of the HQC
set-up \cite{fujii2,pach2}, techniques for the calculation of the holonomies \cite{pach2,marg},
and its intrinsic fault-tolerance \cite{pach3,preskill}.

The methodology is briefly described as follows.  The quantum code
is realized by the $n$-dimensional eigenspace, ${\cal C}$, of an
$n$-fold degenerate Hamiltonian $H_0$ with eigenvalue $E_0$.   Let
$M$ be a $d$-dimensional real parameter space. We suppose that the
experimentalist can implement unitary transformations ${\cal
U}(\eta)$ depending continuously on the parameter $\eta \, \in M
.$  A set of isospectral Hamiltonians is formed by the adjoint
orbit of $H_0,$
\begin{equation}
{\cal O} (H_0) \equiv \{ {\cal U} (\eta)  \, H_0  \, {\cal
U}^{\dag}(\eta) \, , \, \eta \in M \}.
\end{equation}

Let $\gamma : [0,T] \to M$ be a closed curve
in parameter space.  If we traverse this loop sufficiently slowly, the adiabatic theorem
ensures that no energy level crossing will occur and ${\cal O} (H_0)$ forms a family
of Hamiltonians that drive the dynamics.  Let $ | \psi  \rangle_i \in \cal{C}$
be the initial state of the system, then after completing an adiabatic loop in parameter
space, the initial and final states are related by
\begin{equation}\label{gate}
| \psi  \rangle_f = e^{iE_0 T}\Gamma_{A}(\gamma)  | \psi  \rangle_i \,\, \in  \, \, \cal{C}
\end{equation}

\noindent where $e^{iE_0 T}$ is the dynamical phase which will be
omitted in the following by setting $E_0 = 0$.  The matrix,
$\Gamma_A(\gamma) \, \in \, U(n) , $ is the {\it holonomy}
associated to the loop $ \gamma $ and for $n > 1 $, these
non-Abelian holonomies are the logic gates of the quantum
computer. The holonomy or {\it geometric phase} \cite{shap}
depends only on the geometry of the loop $\gamma$ and can be
expressed as
\begin{equation}
\Gamma_A(\gamma) = {\bf P} \mathrm{exp} \oint_{\gamma} A
\end{equation}
\noindent where $\bf{P}$ denotes path ordering and the skew-symmetric matrix valued one-form
$\cal{A}$ is known as the Wilczek-Zee connection \cite{wil} with matrix elements
\begin{equation} \label{WZ}
A_{\eta_i}^{\bar{\varrho} \varrho} \equiv \langle \bar{\varrho} |
{\cal U}^\dag (\eta) \frac{\partial}{\partial \eta_i} {\cal U}
(\eta) | \varrho \rangle.
\end{equation}

In this way, $\Gamma_A$ may be considered to be a map from the
loop space of $M$ to the matrix Lie group $U(n)$.  The set Hol$(A)
$:= $\{ \Gamma_A(\gamma) \, | \, \gamma \in M \} $ forms a group
under composition of loops in parameter space and is, in general,
a \emph{subgroup} of $U(n)$.  Hol$(A)$ is said to be the {\it
holonomy group}, and the corresponding Lie algebra is known as the
{\it holonomy algebra}.   When Hol$(A)$ = $U(n)$, the connection
$A$ is {\it irreducible} \cite{zan}. Clearly, irreducibility of
the connection is sufficient for universal computation since any
$n$-level unitary transformation may be applied to the code $|\psi
\rangle \, \in \, \cal{C}$.

\section{Universality}
Since the work of Montgomery \cite{mont1,mont2,mont3,mont4},
mathematicians and engineers have cast certain problems in the
modelling and control of dynamical systems subject to nonholonomic
constraints in the language of gauge theory. These constraints
come in two main varieties.  A cat in free fall experiences a {\it
dynamical} constraint set by the requirement that its angular
momentum remain constant throughout its descent.  An upside down
cat with initial angular momentum zero endeavors to achieve a
rotation and land upright by altering its shape.  When a
mechanical system interacts with its environment, {\it kinematic}
constraints
are often encountered.  For example, kinematic constraints are in
force for a mobile robot with two independently controlled rear
wheels subject to a no-slip constraint against the rolling surface
\cite{kelly}.

Mathematically, the scenario is described by a {\it connection on a 
principal G-bundle} \cite{nak}.  We recall these constructions and provide a 
geometric setting for the remarks made in the introduction.  
A principal $G$-bundle is formed by manifolds $Q$ (total space),
$M$ (base space)  a free Lie group action $ \Phi: G \times Q \to
Q$, and the canonical projection $Q \xrightarrow{\pi} Q/G \equiv M
.$  If $U$ is a neighborhood of M, then $Q$ is locally
diffeomorphic to the product $ U \times G .$ A smooth map $\sigma
: U \xrightarrow{} Q$, such that $ \pi \circ \sigma  = \rm{id}_M$
is called a local section over $U.$

The fiber $\pi^{-1}(p)$ over a point $p \in M$ is identical to the
group orbit and is denoted $G_p.$ Let $ \mathfrak{g}$ denote the
Lie algebra of $G$.  For any element, $ \xi \in \mathfrak{g}$ the
group action $\Phi_{ \rm{exp}(t\xi) } \, q $ defines a curve
though $q \in Q$ . The \emph{infinitesimal generator} $\xi_q$ of
the group action is defined as the tangent vector
\begin{equation}
\xi_q := \frac{d}{dt} \biggr|_{t=0}   \Phi_{ \rm{exp}(t\xi) } \, q .
\end{equation}
The {\it vertical subspace} $V_qQ$ is defined to be the subspace of $T_qQ$ that is tangent
to the fiber $G_p$, by the previous definition we have the identification
$ V_q Q \cong \mathfrak{g}$.

A connection ${\cal A}$ on $Q$ is an Ad-equivariant Lie
algebra-valued one-form ${\cal A}: TQ \to \mathfrak{g}$ such that
${\cal A}(\xi_q) = \xi $ for all $\xi \in \mathfrak{g}.$ The
horizontal space $H_qQ$ is the linear space $H_qQ := \{ X_q \in
T_q Q | {\cal A}_q( X_q ) = 0 \}$.  The local connection form $A$ is defined 
with respect to a local section $ A = \sigma^*({\cal A})~.$  These definitions provide the
splitting
\begin{equation}
T_qQ = V_qQ \oplus H_qQ
\end{equation}
of the tangent vectors into horizontal and vertical components.
Note that
\begin{equation}
V_q Q = {\rm Ker}T_q \pi \quad {\rm and} \quad H_q Q =  {\rm Ker} {\cal A}_q .
\end{equation}

The projection map at a point defines an isomorphism from  the
horizontal space to the tangent space to the base space by  $T_q
\pi : H_q \to T_{ \pi (q)} M $.  Thus a curve $q(t) \in Q$ defines
a curve in the base space by specifying a tangent vector at each point
$\pi(q(t)) = p(t)$. The properties of the connection and the
uniqueness theory of ODE's provide the reverse procedure of
reconstructing a curve in the total space given a curve in the
base space called the \emph{horizontal lift} \cite{nak}.  Denote
the horizontal lift of $X \in TM$ by $X^h$ and the horizontal part
of $Z \in TQ$ as $h Z$.  The horizontal lift of any closed curve
in the base space maps the fiber to itself and corresponds to a group
element $g$ by the automorphism $q_f = \Phi_g q_i . $ Assuming
direct control over the base velocities, we seek a closed curve in
the base space that achieves a desired group translation in the
fiber.

In the case of the cat (robot), the conservation law (no-slip
constraint) defines a connection on a principal bundle with group
action $SO(3) \times Q \to Q  \,\, \, (SE(2) \times Q \to Q).$ The
key point in the modelling and control of these systems is that
the horizontal distribution, defined by the connection, encodes
the constraint information.  Thus a curve $q(t) \in Q $ satisfies
the constraints if its tangent vector $X_{q(t)}$ lies in
$H_{q(t)}Q $ for all $t$.  Such a curve is called horizontal.
Given an initial configuration $q_i$, the feasibility of reaching
a final configuration $q_f$ is then equivalent to the existence of
a horizontal curve joining $q_i$ and $q_f.$ For control systems of
this type, it is natural to define the {\it reachable set} from
$q_i$ as the set of points $q \in Q$ that lie on a horizontal
curve originating at $q_i$.

\vspace{10pt}
\noindent {\it Theorem (Chow):} Suppose Q is connected.  Let
$X_i^h \, , \,\, 1 \leq \, i \, \leq d   $
be a local frame of the horizontal space at $q$.  Then any two points of $Q$
can be joined by a horizontal curve if the iterated Lie brackets
$ [ X_{i_k}^{h} \, , \, [ X_{i_{k-1}}^h, \dots , [ X_{i_2}^h \, , \,  X_{i_1}^h]  \dots ] $
evaluated at  $q \in Q$ span the tangent space $T_q Q$ for all $q$.
\vspace{10pt}

For control systems without drift, this classical theorem \cite{chow} gives a
sufficient condition to determine whether the reachable set is the entire manifold $Q.$
If the Lie brackets defined in the theorem fail to span all of $T_q Q$, the
reachable set may be characterized as follows.  Denote the subspace of $T_q Q$ defined 
in the theorem as
\begin{equation}
\Delta_q =  {\rm span} \{ [ X_{i_k}^{h} \, , \, [ X_{i_{k-1}}^h, \dots ,
[ X_{i_2}^h , \,  X_{i_1}^h] \dots ]\, ; \, 1 \, \leq  \, i_k \, \leq \, d , \,
1  \, \leq \, k \, < \infty  \}
\end{equation}
\noindent The sub-bundle $\Delta = \bigcup_q \Delta_q$ forms, by
construction, an involutive distribution on the manifold $Q.$  If
the rank of $\Delta_q$ is constant as $q$ is varied, the
Frobenious theorem \cite{isidori, jurd2} then asserts the
existence of an {\it integral submanifold} $\widetilde Q \subset
Q$ with $\Delta$ as its tangent space.  This submanifold is
invariant under the constrained dynamics and forms the reachable
set.

These arguments and the general principle embodied in the theorem
are well known in the quantum computation literature.  In the
usual \emph{dynamical} approach to quantum computing, the
experimentalist has a  repertoire of Hamiltonians $ \{ H_l \}_{l=1}^r
$ that act on the quantum state.  If the Lie algebra generated by
the $H_l$ under commutation is equal to $su(n)$ (or $u(n)$), then
the system is deemed capable of performing universal computation.
This is equivalent to the notion of \emph{complete
controllability} for quantum systems \cite{jurd,tarn,ramakrishna}.
It is not
surprising, then, that Chow's theorem is decisive the 
holonomic
framework as well.

To establish universality of HQC, we must show that the holonomy
group Hol$(A)$ is rich enough to generate a universal set of
logic gates.  The holonomy group is determined by the
$\mathfrak{g}$-valued \emph{curvature two-form} defined by
\begin{equation}
{\cal F} (X_1, X_2) = {\rm {\bf d}} \,{\cal A} ( h X_1 \, , \, h
X_2 )
\end{equation}
where {\bf d} denotes exterior differentiation.  To evaluate the
curvature we employ the structure equation
\begin{equation}
{\cal F} (X_1,X_2) = \mathrm{d} {\cal A} (X_1,X_2) + [ {\cal A} (X_1), {\cal A} (X_2)] \,
\end{equation}
for $X_1, X_2 \in T_q Q$.  Since the connection $\cal{A}$
evaluates to zero on horizontal vectors, the structure equation
implies that $ {\cal F} (X_1^h,X_2^h) = - {\cal
A}([X_1^h,X_2^h]).$ Recalling that a connection can be defined as
a projection of a vector  $ X \in T_q Q $ onto the vertical
subspace $V_qQ \cong \mathfrak{g}$, we see that ${\cal
F}(X^h,Y^h)$ is the vertical component of the vector
$[X_1^h,X_2^h]$. The Ambrose-Singer theorem expresses the holonomy
group associated with the connection in terms of the curvature. As
stated in \cite{nak}:

\vspace{10pt}
\noindent {\it Theorem (Ambrose-Singer):} Let $Q$ be a principal $G$-bundle over a manifold $M$.
The Lie algebra $\mathfrak{h}$ of the holonomy group $\mathrm{Hol}_{q_0}({\cal A})$ of
a point $q_0 \, \in \, Q$ agrees with the subalgebra of $\mathfrak{g}$  spanned by the
elements of the form
${\cal F}_q(X^h,Y^h)$ where $ X^h \, , \, Y^h \, \in H_qQ$ and {\it $q$
is a point on the same horizontal lift as
$q_0$}.
\vspace{10pt}

This theorem has been quoted by other authors to provide sufficient conditions
for universality of HQC.  The statement in italics, however, demands that we
evaluate the curvature on the horizontal space at every point
$q$ that is reachable from $q_0$ via a horizontal curve.  This set of points, however,  is
the reachable set as defined above.

More tractable conditions are obtained from Chow's theorem.
By the above reasoning, elements of the form
\begin{equation}
{\cal F} (X_i^h, X_j^h) = - {\cal A}([X_i^h, X_j^h])\, , \quad  {\cal F} (X_i^h,
X_k^h) = - {\cal A}([X_i^h, X_k^h])\, ,  \,\,\, \dots
\end{equation}
contribute a set of group directions obtained from brackets of horizontal vectors.
According to Chow's theorem we must compute all the {\it iterated} Lie brackets
of horizontal vectors.  The vertical component of the vector $[X_1^h, [X_2^h, X_3^h]]$
is given by $\mathrm{D}_{X_1^h} {\cal F} (X_2^h, X_3^h)$ and higher order Lie brackets
are expressed as higher order covariant derivatives of the curvature.

\vspace{10pt} \noindent {\it Corollary}:  Suppose $Q$ is connected.
The holonomy algebra at a point $q_0 \in Q$ is spanned of by the
curvature forms ${\cal F}(X_{i_{1}}^h,X_{i_{2}}^h)$ and the
covariant derivatives $D_{X_{i_k}^h} \, D_{X_{i_{k-1}}^h} \dots \,
D_{X_{i_3}^h} {\cal F} (X_{i_2}^h , X_{i_1}^h)$ evaluated at $q_0
~. $ \vspace{10pt}

This result appears in Montgomery \cite{mont1}.  A proof can be found in 
\cite{mont4}.  (See also \cite{luc}.) It is interesting
to note that Montgomery's original  motivation, in addition to the cat's
problem, was the optimal control of spin systems.

To apply this result to HQC we identify the relevant manifolds and the horizontal direction.
Following Fujii \cite{fujii1,fujii2},
let ${\cal H}$ be a separable Hilbert space and define the manifolds
\begin{align}
\mathrm{St}_n ({\cal{H}}) &:= \{ V = (v_1, \, \dots \, , v_n) \in {\cal{H}} \times \, \dots \,
\times {\cal{H}} \,\, | \,\, V^{\dag}V = \mathrm{Id}_{ n \times n} \}  \notag \\
\mathrm{Gr}_n ({\cal{H}}) &:= \{ X \in B({\cal H}) \,\, | \,\, X^2=X \, , \, X^{\dag} = X \, , \,
\mathrm{tr}X = n \} \notag
\end{align}
\noindent where $B({\cal H})$ denotes the set of bounded linear
operators on ${\cal H}$. These manifolds are known as the Stiefel
and Grassmann manifolds respectively. They form a principal bundle
with the (right) $U(n)$ action on $\mathrm{St}_n({\cal H})$ and
the projection $ \pi: \mathrm{St}_n({\cal H}) \to
\mathrm{Gr}_n({\cal H})$ given by $\pi(V) = V V^{\dag}$.  Denote
this $U(n)$-bundle by $P_n$. Let $M$ be the parameter space and
let the map $\Pi \, : M \, \to \, \mathrm{Gr}_n({\cal H})$ be
given. The principal bundle of interest is then formed by the
pullback of $P$ by $\Pi$, $Q = \Pi^* P$ with total space
$$
Q = \{ (\lambda, V) \, \in \, M \times \mathrm{St}_n({\cal H}) \, | \, \Pi(\lambda) = \pi(V) \}
$$
over the base manifold $M$.  To be precise, the {\it left} action
of the matrix acting on a vector $ | \psi \rangle \, \in {\bf
C}^n$ as defined by (\ref{gate}) takes place in the ${\bf C}^n$
vector bundle associated to $Q$ \cite{fujii1,fujii2}.

An important special case of this construction, known as the ${\bf CP}^n$ model,
has been shown to be {\it generically} irreducible \cite{zan}.  In this case,
${\cal H} = {\bf C}^{n+1} = \{ | \alpha \rangle \}_{\alpha = 1}^{n+1}$ and $H_0$ has
an $n$-dimensional degenerate subspace.  The parameter space $M = {\bf CP}^n $ is isomorphic
to the orbit of $H_0$,
$$
{\cal O} (H_0) \cong \frac{U(n+1)}{U(n) \times U(1)} \cong \frac{SU(n+1)}{U(n)} \cong  {\bf CP}^n.
$$
Thus $\Pi$ is a surjective map of $M = {\bf CP}^n$ onto ${\cal O}
(H_0) \cong \mathrm{Gr}_{1,n+1}$. Due to the large parameter
space, this model can be shown to be irreducible by considering
the span of the curvature form only (and not its covariant
derivatives).  Note that this model requires control over $2n =
\mathrm{dim}_{\bf R} {\bf CP}^n $ parameters to control an
$n$-level system.

In any case, the  Wilczek-Zee (\ref{WZ}) connection with its built
in Hermitian structure defines the horizontal subspace by
identifying horizontal vectors as those which are orthogonal to
the fiber \cite{wil,pach2}.  When applied to HQC, this result
represents a significant reduction in the control resources
necessary for universality and thus broadens the class of quantum
evolutions that are capable of computation. Indeed, if one
considers the span of the curvature form only, then one {\bf
incorrectly} concludes that a necessary condition for universality
of an $n$-level system is given by $ d(d-1)/2 \geq n^2$ where $d =
\mathrm{dim}_{\bf R} M $ \cite{pach2}.

\section{Optical Holonomic Computer}
It is widely believed that coherent superposition alone cannot
account for the exponential speed-up sought in the realization of
a quantum computer.  Quantum entanglement must also be present
\cite{josza}. As observed in \cite{zan,pach2} the ${\bf CP}^n$
model does not possess a multi-partite structure necessary for
encoding
entangled states. Attention is therefore directed to physical
systems that have a multi-partite structure built in from the
start {\it and} over which the experimentalist can exert control.
While the results presented here apply to any HQC set-up, we are
interested in a promising model coming from quantum optics where
displacing and squeezing devices realize control operations acting
on laser beams in a non-linear Kerr medium
\cite{zan}-\cite{pach2}.

Let $a^{\dag} \, , \, a$ be creation and annihilation operators of the
harmonic oscillator and let $n = a^{\dag}a$ be the number
operator.  Let $ \cal{H} $ be the Fock space generated by  $a^{\dag} \, \mathrm{and} \, a $
with basis $ \{ | \nu \rangle \, : \, \nu = 0 \, , \, 1 \, \dots \, \, \}$.
Each qubit is encoded in the degenerate subspace of the interaction Hamiltonian
\begin{equation}
H^1 = X \hbar n(n-1)
\end{equation}
where $X$ is a constant \cite{pach1}.   This computing scheme
scales to a system of $m$ qubits by employing $m$ lasers to form
the product basis $ | \nu_1 \nu_2 \dots \nu_m \rangle =   | \nu_1
\rangle  \otimes | \nu_2  \rangle \otimes \dots | \nu_m \rangle$
where $ \nu_i \in \{ 0 , 1 \}$ . In accordance with the quantum
circuit model, a control strategy is devised to implement all
single qubit rotations and a non-trivial two-qubit transformation.
A fundamental result then asserts that universality of the entire
quantum register of $m$ qubits can be achieved by this set of
two-level local transformations \cite{div}.

Single mode squeezing and displacing operators are employed to control the single qubit,
\begin{equation}
S(\mu) \, = \, \mathrm{exp} \, (\mu a^{{\dag}{2}} - \bar{\mu}
a^{2}) \quad \quad D(\lambda) \, = \, \mathrm{exp} \, (\lambda
a^{\dag} - \bar{\lambda}a)
\end{equation}
where $ \mu ~, \lambda ~ \in {\bf C} ~.$  These operators define a
two parameter orbit of $H_1$ under the unitary transformation
${\cal U}(\lambda,\mu) = D(\lambda)S(\mu),$
\begin{equation}
{\cal O}(H^1) = {\cal U}(\lambda,\mu) H^1  {\cal U}^{\dag}(\lambda,\mu) \, .
\end{equation}
The holonomy group associated to loops in the $(\lambda,\mu)$ parameter space is
$U(2)$ \cite{pach1,fujii1}.

To prove universality of the computational model it suffices to
generate non-trivial $U(4)$ holonomies.
For the two-qubit system, the Hamiltonian is given by,
\begin{equation}
H^{12} = X \hbar  n_1(n_1-1) + X \hbar  n_2(n_2-1) ,
\end{equation}
where $n_i$ is the number operator for the $i$-th beam. Two-mode
squeezing and displacing operators realize control operations,
\begin{equation}
 M(\zeta) = \mathrm{exp} \, (\zeta a_1^{\dag}a_2^{\dag} - \bar{\zeta} a_1a_2)
\quad \quad
 N(\xi) = \mathrm{exp} \, (\xi a_1^{\dag}a_2 - \bar{\xi}a_1a_2^{\dag})
\end{equation}
where $ \zeta  = r_2 e^{i \theta_2} \, , \,  \xi = r_3 e^{i
\theta_3} ~ \in {\bf C} ~. $ In the adiabatic limit, the adjoint
orbit under the action ${\cal U}(\xi , \zeta ) = N(\xi)M(\zeta)$,
\begin{equation}
{\cal O}( H^{12} ) = {\cal U}(\xi,\zeta) H^{12}  {\cal
U}^{\dag}(\xi,\zeta)
\end{equation}
drives the dynamics.  The degenerate subspace of $H^{12}$ is given by
the computational basis $ \{ | 0 0 \rangle , | 0 1 \rangle , | 1 0
\rangle, | 1 1 \rangle \}$.  Set $ | \mathrm{vac} \rangle = ( | 0
0 \rangle , | 0 1 \rangle , | 1 0 \rangle, | 1 1 \rangle )  \, \in
\,  \mathrm{St}_4 ({\cal H} \otimes {\cal H}).$

We characterize the reachable set from $q_0 = ( m , | \mathrm{vac}
\rangle ) \in Q $ by applying the conditions obtained from Chow's
theorem. The local connection coefficients $A_\nu$ are written
in terms of the base variables only, $ (r_2, \theta_2, r_3,
\theta_3) \in  M  \subset {\bf C}^2 \,  $ \cite{pach1, pach2},
\begin{gather}
A_{r_{2}} = \left[
\begin{matrix} 0 & 0 & 0 & -e^{-i\theta_2} \\
0 & 0 & 0 & 0 \\ 0 & 0 & 0 & 0 \\ e^{i\theta_2}& 0 & 0 & 0
\end{matrix}
\, , 
\right]  \quad A_{r_{3}} = \left[
\begin{matrix} 0 & 0 & 0 & 0 \\ 0 & 0 & -e^{-i\theta_3} & 0 \\
0 & e^{i\theta_3} & 0 & 0 \\ 0 & 0 & 0 & 0 \\
\end{matrix}
\right] (2\cosh^2r_2 - 1)  \, , \notag \\
A_{\theta{_2}}  \, = \, \left[
\begin{matrix}
0 & 0 & 0 & e^{-i\theta_2} \\
0 & 0 & 0 & 0 \\0 & 0 & 0 & 0 \\ e^{i\theta_2}& 0 & 0 & 0
\end{matrix}
\right] \frac{i}{2} \sinh2r_2 \, + \, \left[
\begin{matrix}
1 & 0 & 0 & 0 \\
0 & 2 & 0 & 0 \\ 0 & 0 & 2 & 0 \\ 0 & 0 & 0 & 3
\end{matrix}
\right]  \frac{i}{2} ( \cosh2r_2 - 1 ) \, ,   \notag \\
A_{\theta{_3}} \, = \, \left[
\begin{matrix}
0 & 0 & 0 & 0 \\ 0 & 0 & e^{-i\theta_{3}} & 0 \\
0 & e^{i\theta{3}} & 0 & 0 \\ 0 & 0 & 0 & 0
\end{matrix}
\right] \frac{i}{2} \cosh 2r_2 \sin 2r_3 \, + \, \left[
\begin{matrix}
0 & 0 & 0 & 0 \\ 0 & 1 & 0 & 0 \\ 0 & 0 & -1 & 0 \\ 0 & 0 & 0 & 0
\end{matrix}
\right] i \sin^2 r_3   \, . \notag
\end{gather}

\noindent The non-zero local curvature forms  $F_{\mu \nu} \, ,$
\begin{gather}
F_{r_{2}r_{3}} \, =  \, \left[
\begin{matrix} 0 & 0 & 0 & 0 \\
0 & 0 & -e^{-i\theta_3} & 0 \\ 0 & e^{i\theta_3} & 0 & 0 \\ 0& 0 &
0 & 0
\end{matrix}
\right]  2 \, \sinh 2r_2   \,, \,  \, \, \, 
F_{r_{2}\theta_{2}} \, =
\, \left[
\begin{matrix} 0 & 0 & 0 & 0 \\ 0 & 1 & 0 &  0 \\
0 &  0 & 1 & 0 \\ 0 & 0 & 0 & 2 \\
\end{matrix}
\right]
2 i \,  \sinh 2r_2 \, ,   \notag \\
F_{r_{2}\theta_{3}}   \, =  \, \left[
\begin{matrix}
0 & 0 & 0 & 0 \\
0 & 0 & e^{-i \theta_{3}} & 0 \\ 0 & e^{i \theta_3}& 0 & 0 \\ 0& 0 & 0 & 0 \\
\end{matrix}
\right]  \, i \sin 2r_3 \sinh 2r_2  \, ,  \, \, \, \notag
F_{r_{3}\theta_{3}} \, =  \, \left[
\begin{matrix} 0 & 0 & 0 & 0 \\ 0 & -1 & 0 &  0 \\
0 &  0 & 1 & 0 \\ 0 & 0 & 0 & 0 \\
\end{matrix}
\right]
 i \,  \sin 2 r_3 \sinh^2 2r_2 \, ,  \notag
\end{gather}
\noindent span  $ su(2) \times u(1) ~. $

The block structure of these matrices suggests that new group directions
may be obtained by taking covariant derivatives of $F_{r_2 \theta_2}$
along the base coordinate vectors $\frac{\partial}{\partial \theta_2} $ and
$\frac{\partial}{\partial r_2} $ ,
\begin{gather}
D_{\frac{\partial}{\partial \theta_2}} F_{r_{2}\theta_{2}} \, =
\, \left[
\begin{matrix}
0 & 0 & 0 &  - e^{-i\theta_2} \\ 0 & 0 & 0 & 0 \\
0 & 0 & 0 & 0 \\ e^{i \theta_2} & 0 & 0 & 0
\end{matrix}
\right]
 2 \sinh^2 2 r_2   \notag \\
D_{\frac{\partial}{\partial r_2}} F_{r_{2}\theta_{2}} \, =  \,
\left[
\begin{matrix}
0 & 0 & 0 &  e^{-i\theta_2} \\ 0 & 0 & 0 & 0 \\
0 & 0 & 0 & 0 \\ e^{i \theta_2} & 0 & 0 & 0
\end{matrix}
\right]-4i\sinh 2r_2 \, + \, \left[
\begin{matrix} 0 & 0 & 0 & 0 \\ 0 & 1 & 0 &  0 \\
0 &  0 & 1 & 0 \\ 0 & 0 & 0 & 2 \\
\end{matrix}
\right] 4i \cosh 2r_2 \, ,  \notag \\
 D_{\frac{\partial}{\partial
\theta_2}} D_{\frac{\partial}{\partial \theta_2}}
F_{r_{2}\theta_{2}} \, =  \, \left[
\begin{matrix}
1 & 0 & 0 & 0 \\ 0 & 0 & 0 & 0 \\ 0 & 0 & 0 & 0 \\ 0 & 0 & 0 & -1
\end{matrix}
\right] 2 i \sinh^3 r_2 \, + \, \left[
\begin{matrix}
0 & 0 & 0 & e^{-i\theta_2} \\ 0 & 0 & 0 & 0 \\
0 & 0 & 0 & 0 \\ e^{i \theta_2} & 0 & 0 & 0
\end{matrix}
\right]  \, 2 i \sinh^2 2r_2 \cosh 2r_2   \notag
\end{gather}

\noindent These matrices and the independent contributions of the previous set span the Lie algebra
$su(2) \times su(2)  \times u(1) \,\, \subset  \,\, u(4)$.  The connection is not irreducible,
however, the product structure of the subgroup shows that non-trivial $U(4)$ transformations
are attainable.  This result reconciles conflicting results from the literature.  One concludes after
consideration of the curvature form only the holonomy group to be
$SU(2) \times U(1)$ \cite{fujii2}.  However, a variant of the \emph{square root of SWAP} gate,
\begin{equation}
U =  \frac{1}{\sqrt{2}}
\left[
\begin{matrix}
\sqrt{2} & 0 & 0 & 0 \\ 0 & 1  &  - i  & 0 \\
0 &  - i  &  1  & 0  \\ 0 & 0 & 0 & \sqrt{2}
\end{matrix}
\right] \,
\end{equation}
can been explicitly constructed \cite{pach2}.  This universal transformation is an
element of $SU(2) \times SU(2)$.

\section{Concluding Remarks}

Holonomic Quantum Computation represents a novel approach to quantum
computing by employing non-Abelian geometric phases to
perform information processing.  The geometric phase is a beautiful
phenomena with a long history in physics, mathematics, and engineering.
We have shown that a result from control theory provides key insight
into the foundations of HQC and opens up the
possibility that more physical systems will be amenable to this approach.
For a particular manifestation of HQC -  the Optical Holonomic Computer -
the results presented here provide
new understanding of the controlled interactions attainable in the
experimental set-up.  Moreover, we have proven the existence of a universal
set of logic gates.

\vspace{10pt}
\noindent {\it Acknowledgements  }:
I would like to thank Fabio Celani, Claudio DePersis, and Paolo Rinaldi
for helpful discussions and Professor Richard Montgomery for introducing
me to the subject.

\end{document}